# Photonic Higher-Order Topological States Induced by Long Range Interactions

Mengyao Li[1,2], Dmitry Zhirihin[1,3], Dmitry Filonov[3], Xiang Ni[1,2], Alexey Slobozhanyuk[3], Andrea Alù[1,2,4], Alexander B. Khanikaev[1,2,4]

[1]Department of Electrical Engineering, Grove School of Engineering, City College of New York, NY 10031, USA.
[2]Physics Program, Graduate Center of the City University of New York, New York, NY 10016, USA.
[3]Department of Nanophotonics and Metamaterials, ITMO University, St. Petersburg 197101, Russia.
[4]Photonics Initiative, Advanced Science Research Center, City University of New York, New York, NY 10031, USA.

**Abstract:** The discovery of topological phases has recently led to a paradigm shift in condensed matter physics, and facilitated breakthroughs in engineered photonics and acoustic metamaterials. Topological insulators (TIs) enable the generation of electronic, photonic, and acoustic modes exhibiting wave propagation that is resilient to disorder, irrespective of manufacturing precision or unpredictable defects induced by the operational environment, known as topological protection. While originally limited to a dimensionality of the protected states that is *one* dimension lower than the host TI material, the recent discovery of higher-order topological insulators (HOTIs) provides the potential to overcome this dimensionality limitations by offering topological protection over an extended range of dimensionalities. Here we demonstrate 2D photonic HOTI (PHOTI) with topological states *two* dimensions lower than the one of the host system. We consider a photonic metacrystal of distorted Kagome lattice geometry that exhibits topological bulk polarization, leading to the emergence of 1D topological edge states and of higher order 0D states confined to the corners of the structure. Interestingly, in addition to corner states due to the nearest neighbour interactions and protected by generalized chiral symmetry [1], we discover and take advantage of a new class of topological corner states sustained by long-range interactions, available in wave-based systems, such as in photonics. Our findings demonstrate that photonic HOTIs possess richer physics compared to their condensed matter counterparts, offering opportunities for engineering novel designer electromagnetic states with unique topological robustness.

**Main text:** Topological systems exhibit unique and often counterintuitive properties, such as robust electronic transport and wave propagation, which promise to revolutionizing technologies across different fields, from quantum electronics [2,3,4,5,6,7,8] to photonics [9,10,11,12,13,14,15,16,17,18,19,20,21,22,23,24,25,26] and acoustics [27,28,29,30,31]. In electronics and quantum photonics, topological phenomena unlocked novel approaches for quantum computing interfaces [32,33] and robust lasing [34,35,36], while in a broad range of classical fields, including optics, mechanics and acoustics, they offer an unprecedented degree of control via synthetic degrees of freedom and robustness, which manifests itself as resilience to defects and disorder [11,12,13,14,15,16,17,18,19,20,21,23,25,27,29,30,31,37]. While most topological systems studied so far have been characterized by the presence of topological states with dimensionality *one* order lower than the one of the system, recently a new class of topological systems, so called higher-order topological insulators (HOTIs), have been introduced [38,39,40,41]. As opposed to conventional topological insulators, HOTIs support topological states [39,40,42,43,44,45,46,47] two and more dimensions lower than the system itself, referred to as higher-order topological (HOT) states. One example of such systems is given by quadrupole topological insulators [38,48], which have been recently implemented in mechanical [49] and photonic [50, 51] systems, as well as in electrical circuits [52]. Another class of crystalline insulators with zero Chern number is the one of Wannier-type second- and third-order topological insulators, which exhibit topological bulk polarization and support zero-energy states localized at the corners [1, 42, 53, 54, 55, 56, 57]. Topological corner states in such systems appear to be protected by time reversal and/or spatial symmetries [41, 53], which may not only pin them to zero energy and localize at one of the sublattices, but also gives rise to their nonradioactive character and a bound states in the continuum behaviour [1].

Here, we design and experimentally demonstrate photonic higher-order corner states in a microwave metacrystal with topological bulk polarization. In addition to higher-order states consistent with those observed recently in acoustics [1], we show that in electromagnetic systems where far-field interactions among non-neighbouring unit cells are inevitable, the coupling beyond nearest neighbours leads to the emergence of a new class of higher order topological corner states, which split from the edge states continuum. Our findings therefore open new opportunities for topological photonic metamaterials, beyond previously considered approximations based on analogies with electronic systems, and envision devices whose functionality relies on multiplicity of topological corner states of different nature interacting with each other and with edge states in a controllable manner.

We explore a two-dimensional photonic topological metacrystal [42, 58] with topological bulk polarization, supporting a topological phase protected by lattice symmetries. The lattice represents an array of dielectric cylinders arranged to form kagome lattice between two parallel copper plates, as shown in Fig.1**a**. Dielectric cylinders are formed by filling the patterned Styrofoam mould with a high index dielectric powder ($n=3.1$). In this geometry, the dielectric cylinders support a vertical dipolar mode that is also the lowest frequency mode we choose to work with to design the topological metacrystal. In addition to near-field coupling of the dipolar modes in the lattice, the modes also couple to the transverse electromagnetic mode supported by the parallel plate host waveguide formed by aluminium plates, thus giving rise to far-field radiative coupling. The latter coupling renders previous analytical treatments of topological Kagome lattices inadequate for the photonic system at hand, as we need to consider interactions beyond nearest neighbour approximations (Supplementary Section 1) and apply first principles methods to rigorously analyse the system. Nonetheless, the coupling strength between cylinders can still be tuned by shifting the cylinders closer to or farther away from one another, thus enabling control over the coupling within the lattice. By comparing the analytical tight-binding model (TBM) and first-principles simulations (using the finite-element-method (FEM) software COMSOL Multiphysics and the Radio Frequency module), we find that the system is well approximated by the tight-binding model (TBM) when next nearest-neighbour coupling is taken into account (Supplementary Section 1).

For the case of an ideal (unperturbed) Kagome lattice, for which the distance between cylinders in the same and nearby unit cells is equal (Fig.1 **b** inset, position 1), i.e., the inter-cell and intra-cell couplings are the same, the band diagram obtained with both TBM (Supplementary Section 1) and FEM simulations (blue solid lines in Fig. 1**b**) exhibits a Dirac-like degeneracy at the K (and K′) point in the Brillouin zone. The degeneracy is formed between low-frequency monopolar modes, characterized by in-phase orientation of the field in all three cylinders of the trimer, and dipolar modes, which are left- and right-handed circularly polarized at the K and K′ points, respectively.

When the metacrystal symmetry is reduced by shifting the dielectric cylinders as shown in Fig. 1**b** inset (positions 2 and 3), the inter-cell and intra-cell couplings between neighbouring trimers are no longer equal, we induce a topological transition manifested in the opening of a complete photonic band gap. The topologically nontrivial expanded structure (position 3) and trivial shrunken structure (position 2) have identical band structures, shown in Fig. 1b by red solid lines. Therefore, the two systems cannot be distinguished by frequency dispersion alone. However, these two scenarios have two distinct topological phases, separated by the gapless (position 1) transition point corresponding to the ideal Kagome lattice. The topological transition appears due to the symmetry reduction from six-fold ($C_6$) to three-fold rotational symmetry ($C_3$), giving rise to the hybridization and band inversion of formerly degenerate dipolar and monopolar bands [59]. The topological transition can be demonstrated directly by calculating the topological bulk polarization through a Wilson loop, and also by investigating the $C_3$-related properties at high-symmetry points in the Brillouin zone, as discussed in the Supplement S2. Given the bulk-boundary correspondence principle for our system [1], while the shrunken and expanded geometries have identical band diagrams, the response in finite-size systems is expected to be very different, as the topologically non-trivial geometry is expected to support boundary protected modes.

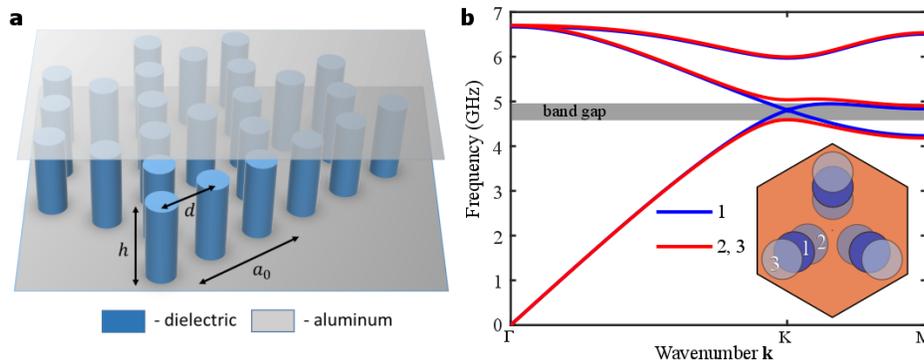

Fig. 1| Kagome lattice and corresponding band spectra. **a** Kagome lattice with dielectric pillars as sites, applying perfectly conducting boundary condition at top and bottom using aluminium plates. **b** Band spectra of unperturbed (1), shrunken (2) and expanded (3) Kagome lattices. Complete bandgap formed in shrunken and expanded case due to partial symmetry breaking. Corresponding lattice unit cells are shown in **b** inset.

The topological bulk polarization characterizes the displacement of the average position of Wannier states from the centre of the cylinders within the unit cell [40, 48]. In the trivial case, the zero value of bulk polarization implies that Wannier states are pinned to the center of the cylinders, and no modes can dangle at the open boundaries or at the domain walls. However, in the topological nontrivial case, the non-vanishing bulk polarization induces a shift of the Wannier states, so that, when the boundaries are introduced, dangling states emerge at the boundaries. A detailed theory describing the topological transition between shrunken and expanded structures can be found in [1]. The bulk polarization difference of the expanded Kagome lattice acquires values (1/3,1/3) relative to the shrunken lattice, leading to the emergence of edge and corner states in crystals possessing interfaces between the two topologically distinct structures.

The nontrivial polarization difference manifests itself in the emergence of topological edge states, which appear at domain walls separating shrunken and expanded metacrystals. These states can be clearly seen in the eigenfrequency simulations of a supercell of 20 unit cells with domain wall in the centre, and at the outer edges (due to periodic boundary conditions imposed at the outer vertical boundaries). The field profiles in Fig. 1c show two modes localized at the inner and outer interfaces, respectively, and the corresponding bands are clearly visible in the band diagram, Fig. 1c (blue solid lines), within the topological band gap separating the bulk continuum (red shaded regions).

In addition to the edge states (in Fig. 2**a**,**b**) in a periodic supercell, the results of first-principle FEM simulations of the finite triangular-shaped structure reveal the presence of zero-dimensional (0D) states localized to the corners of finite structures (Fig. 2**c**), along with edge states confined to the boundaries between trivial and nontrivial domains (Fig. 2**d**). The corner states in Fig. 2**c**, which will be referred to as type I corner states, are analogous to those predicted in Kagome lattices with near-neighbour coupling [42] in condensed matter systems and to their counterparts recently observed in acoustics [1, 57]. These states represent a class of higher-order (D-2) topological states, which are confined to corners of the system when the angle of the corner equals 60°. In the case in which the nearest-neighbour coupling dominates, these states appear to be pinned to a single frequency, referred to as zero-energy, which corresponds to the resonance frequency of a single isolated cylinder.

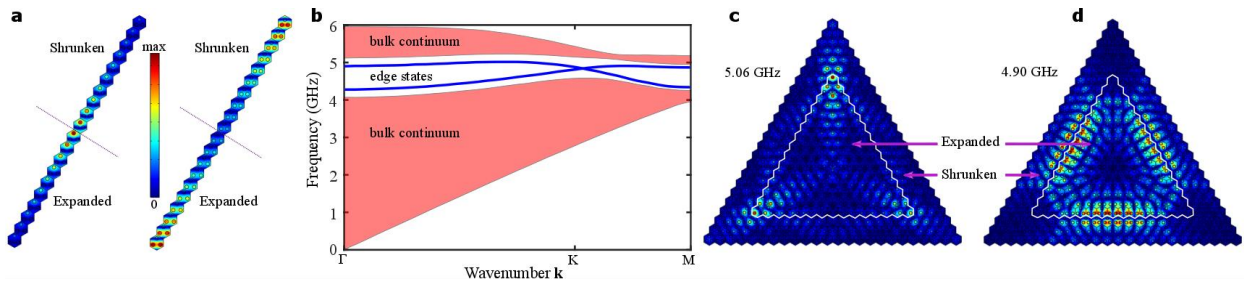

Fig. 2| Simulations of edge and corner states at nontrivial (expanded) and trivial (shrunken) lattice boundary. **a** Kagome supercell structure with shrunken/expanded domain walls, and **b** the corresponding band spectrum. **c** and **d** shows the corner and edge states in finite, triangle shaped array.

These type I 'zero-energy' corner modes are protected by the generalized chiral symmetry $\Gamma_3$ [1] specific to the structure with three sublattices lacking interactions within the same sublattice. In the electromagnetic scenario, however, the interaction between cylinders in the same sublattice cannot be neglected, although it is expected to be smaller than the one between neighbouring cylinders (which belong to different sublattices), due to the drop of coupling with distance. As shown below, these interactions break the generalized chiral symmetry $\Gamma_3$, leading to a correction in the frequency of type I HOT corner states.

Another important distinction between electromagnetic and other topological system resides in the boundary conditions that may terminate a finite system. The boundary conditions can have detrimental effects on corner or edge states, e.g., they can break the symmetry responsible for the very emergence of these states, thus destroying them. In addition, open boundary conditions and coupling to the radiation continuum may give raise to radiative decay of the boundary modes, leading to finite lifetime and propagation length. To suppress such radiation channels, and verify the robustness of our system to perfect electric conductor boundaries, we numerically calculated the eigenstates of the expanded triangular system in Fig. 2**c**,**d**. We find that both corner and edge states are robust, and remain highly confined to the edges and corners of the structure exhibiting only marginal eigenfrequency shift.

The results of first principles simulations for the crystal with perfectly conducting enclosure (Fig. 3**a**) show the evolution of the band structure of the finite metacrystal as the system undergoes a topological transition when the ratio of inter- to intra-cell distance is gradually detuned. As expected from the above discussion, the topological states appear only in the nontrivial case of the expanded lattice, characterized by non-vanishing bulk polarization. One can also clearly see the effect of broken generalized chiral symmetry due to the long-range interactions in the crystal; instead of being pinned to a single frequency (zero-energy) [1, 60], the spectral position of type I corner states shifts slightly towards lower frequencies. Note also the existence of a small bulk bandgap in the spectrum at the topological transition point ($\kappa = 2d/a_0 = 1$) in Fig 3**a** due to the finite size of the lattice. We numerically verified that this gap closes in the limit of very large structures.

Another interesting observation can be made by inspecting Fig. 3**a** and corresponding eigenmodes field profiles Fig. 3**b-d**. Three modes split off from the bulk continuum and follow closely the edge states spectrum as the coupling detuning parameter $\kappa$ enters the topological regime. For values of $\kappa$ slightly exceeding unity, the corresponding eigenmodes have a field profile similar to the one of edge states, but they tend to localize to corners as $\kappa$ increases, establishing an interesting new type II of HOT corner states. Interestingly, the conventional tight binding model, which considered only nearest-neighbour coupling fails to predict these states, thus suggesting that their nature is specific to wave-based systems, and is linked to the interactions beyond nearest-neighbouring cylinders. Indeed, by including next-nearest-neighbour interactions within the tight binding model reveals this new type II corner modes, which, just as in the case of first-principles electromagnetic studies, split off from the edge states (Fig. S1 in Supplement section S1) and have field profiles similar to that of the edge states, but are exponentially decaying away from the corners. This observation allows us to draw the conclusion that type II HOT states appear due to localization of topological edge states at the corners caused by long-range interactions (more details are available in Supplement section S1). Our numerical results also prove that such long-range interactions (beyond nearest neighbour) preserve quantized bulk polarization of (1/3,1/3) (Supplement S2).

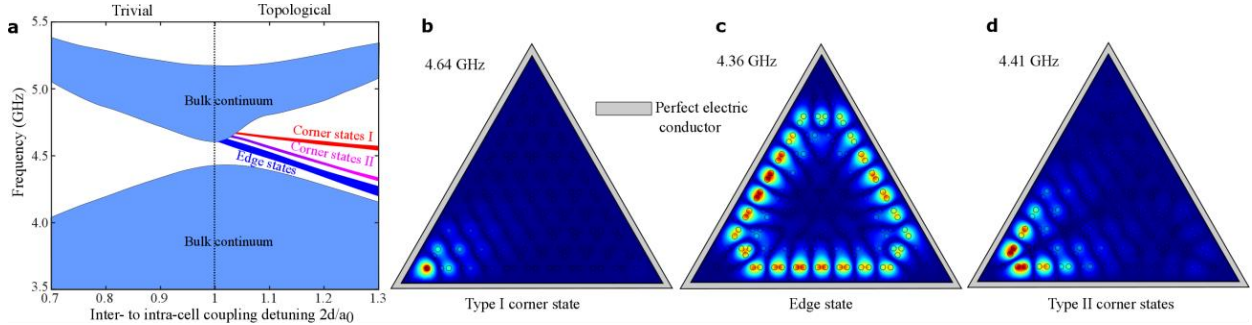

Fig. 3| Simulation results of edge states and corner states type I&II with perfect electric conductor (PEC) boundary condition. **a** Energy spectrum for triangle Kagome lattice with coupling detuning parameter $\kappa = 2d/a_0$. **b**, **c** and **d** showing the field profiles and their corresponding frequencies. Both corner states (type I as red band and type II as magenta band in **a**) and edge states (dark blue band in **a**) appear in band gap of topological region and are gapped from each other. Corner states I and II are both triply degenerated states.

In order to experimentally validate our theoretical predictions, we have fabricated a metacrystal with 10 unit cells per triangle length and performed near-field measurements in a parallel-plate waveguide. The metacrystal was realized in a Styrofoam triangular substrate (permittivity around 1.1 and negligible losses) with machine-processed holes filled by commercially available dielectric powder (Eccostock HiK with the permittivity ε = 10 and loss tangent tan δ = 0.0007). The thickness of the substrate was 20 mm. The fabricated metacrystal was wrapped by conducting (aluminium) foil and the modes were excited at the corner by placing a dipole antenna into the hole drilled in the lower metal plate. The field probing was done by another dipole antenna placed into another hole drilled in the upper plate, with both plate and antenna moving together across the structure to map the field distribution. Fig. 4a presents the extracted local density of states, retrieved from the measured field profiles by applying different filter functions that take into account the field profiles of the respective modes. For type I corner states, the filter is the Heaviside step functions placed over the three corner cylinders. For type II corner states, the filter is the step functions over pairs of cylinders adjacent to the corner one. For the edge states, all edge cylinders except those at and adjacent to the corner are included into the filter, while for bulk states, all internal cylinders are considered. As seen from markers (dots) extracted from first principles eigenfrequency studies, the extracted experimental densities of states are in excellent agreement with the results of theoretical calculations as they all peak at the respective frequencies of the type I and II corner states, and also overlap spectrally the edge continuum. The field profiles corresponding to the frequencies of the maximal density of corner states are also consistent with the theoretical results in Fig. 3**a**,**c**. In particular, for type I states both theoretical and experimental results clearly reveal localization of the field to the sublattice former by lower left cylinders, with only marginal penetration to nearby cylinders of the other sublattices. This demonstrates that, despite the fact that the sublattice symmetry is broken by the interactions within the same sublattice, and the interactions in general have a long range character, the degree of symmetry reduction is not sufficient to destroy the features of type I corner states. Moreover, the interactions beyond nearest neighbours clearly give rise to the emergence of a new type II topological higher order state whose field profile (Fig. 3**c**) confirms its origin from edge continuum, localized due to such interactions. (Note that the bright spot at the corner is the result of the exciting dipole placed at the corner.)

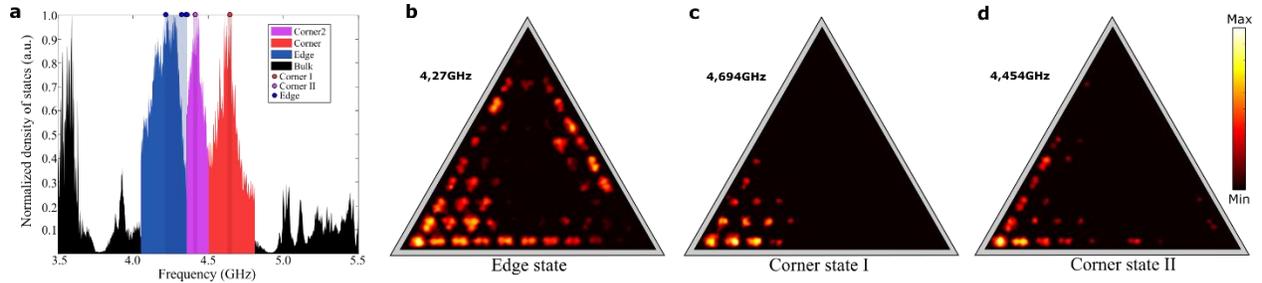

Fig. 4| Experimental observation of HOT Type I and Type II corner states, and density of states at corner, edge and bulk. **a** Density of states experimental results and comparison to eigenstate simulations. Coloured areas are experimental results of corner excitation, and corresponding coloured dots at the top of the figure are first principle calculations of the eigenvalues for the same structure. **b**, **c** and **d** different types of corner states and edge state observed in experiment at corresponding frequencies when the system is excited by a source placed at the left bottom corner, log scaled field profiles from experiment are shown in these figures.

To conclude, in the present work we have demonstrated that the physics of higher order topological states in photonic structures can be quite distinct from their condensed matter and acoustic counterparts, despite a similar geometry and topological phase. In particular, the presence of near- and far-field interactions in electromagnetic systems provides richer opportunities to break symmetries, responsible for the protection of higher-order states, in turn enabling topological states not available in electronic systems. In the Kagome metacrystal considered here, we found that such interactions give rise to a spectral shift of type I corner states, without destroying them (e.g. by merging it with the bulk continuum or with the edge spectrum). More importantly, we found theoretically and confirmed experimentally that long-range interactions can result in the formation of a new class of higher order states - type II corner states. Our studies also reveal an inherent robustness of higher-order states to different boundary conditions, making PHOTIs suitable for integration into different device platforms. In particular, we have found that corner states persist when the crystal is terminated by conducting boundary conditions, and confirmed this robustness experimentally.

Our study demonstrates significant potential of PHOTIs for revealing new fundamental phenomena stemming from physics that is specific to electromagnetic systems. It also shows that PHOTIs represent an exceptional platform for the realization of tunable and robust devices with highly controllable field localization and spectral characteristics. The peculiar interplay between long-range interactions and topology has been shown to lead to the transformation of edge states into a new class of corner states, which enables coupling topological states of different dimensionalities. Extension of our ideas to higher-dimensional photonic systems, and to synthetic dimensions, opens even broader opportunities for novel PHOTI states. Originating from lower-order topological states due to their localization induced by long range interactions, such states can be tailored on demand to design and construct a variety of interfaces and platforms of interacting topological states of different orders.

**Acknowledgements**
The work was supported by the Defense Advanced Research Projects Agency under the Nascent programme with grant number HR00111820040, by the National Science Foundation with grant numbers EFRI-1641069 and DMR-1809915.

**Data availability**

Data that are not already included in the paper and/or in the Supplementary Information are available on request from the authors.

**Author contributions**

All authors contributed extensively to the work presented in this paper.

**Competing interests**
The authors declare no competing interests.

**Corresponding authors**
Correspondence to Alexander B. Khanikaev.